# A Distributional Representation Model For Collaborative Filtering

Zhang Junlin, Cai Heng, Huang Tongwen, Xue Huiping
Chanjet.com
{zhangjlh,caiheng,huangtw,xuehp}@chanjet.com

## Abstract

In this paper, we propose a very concise deep learning approach for collaborative filtering that jointly models distributional representation for users and items. The proposed framework obtains better performance when compared against current state-of-art algorithms and that made the distributional representation model a promising direction for further research in the collaborative filtering.

## 1. Introduction

Recommender systems are best known for their usage on e-commerce websites. By bringing much more extra profit for the website, better recommendation algorithms have attracted attention both from the industry and the academic community. Collaborative Filtering (CF) is one of the most popular approaches among the recommendation algorithms. It utilizes user feedback to infer relations between users, between items, and ultimately relate users to items they like. On the other side, recent years have witnessed the breakthrough of applying the deep learning algorithms into the object recognition[1][2] and speech recognition[3][4]. The NLP is another filed in which the deep learning is widely used. Inspired by successful application of deep learning in NLP tasks [5][6][7][8][9], especially the distributional representation method, we want to explore the distributional representation of users and items for collaborative filtering.

In this paper, we proposed the framework which combines the three-layer neutral network with the distributional representation of the users and items for collaborative filtering. By explicitly encoding the features into vectors, we can explore the complex nonlinearity interdependencies of features through this neutral network. Though seems to be simple, the method has been proved to be effective in recommendation domain by experiment results.

The main contributions of this work can be summarized as following:

We propose a distributional representation approach for recommender system which, to the best of our knowledge, is the first study to introduce the word embedding concept into collaborative filtering. The experiment results show that it's a promising direction for further research.

Section 2 describes the distributional representation framework for collaborative filtering. In Section 3, we present the experiment results which indicate the proposed method outperforms many commonly used algorithms in this research field. Section 4 presents a brief overview of related work. The Final section is the conclusion of this paper.

## 2. Distributional Representation Model For Recommendation

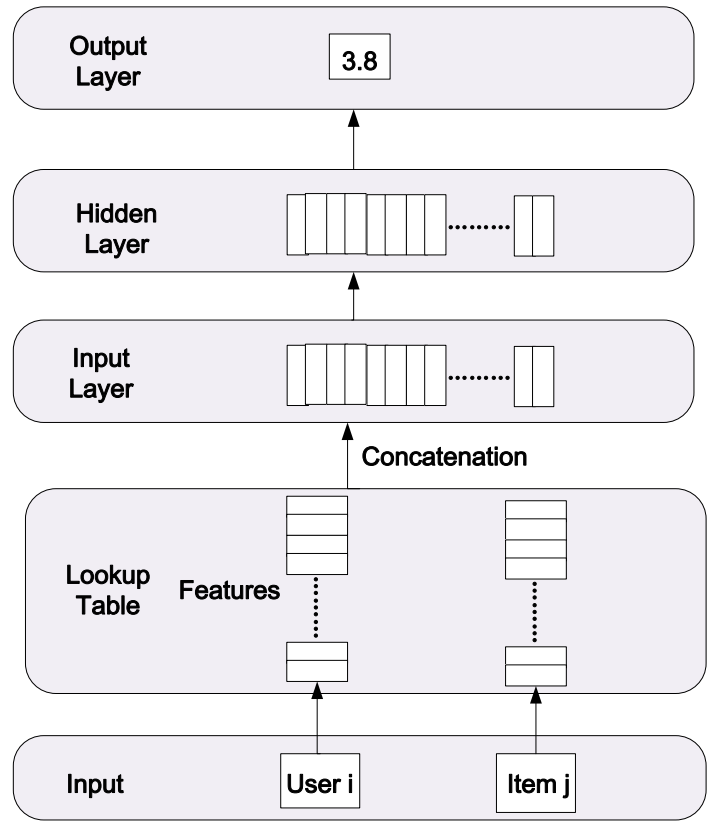

Fig 1.　Neutral network structure of distributional representation model

Collaborative Filtering is one of the most popular approaches for building recommendation systems. CF mostly relies on past user's behavior such as their previous transactions or product ratings (For convenience, we will call the transaction or product as item in the following part of this paper). In order to identify new user-item associations, It analyzes relationships between users and interdependencies among items. Our proposed model explicitly transforms the user and item into vectors which encode the latent features and then it uses the combined vectors as neutral network's input to explore the complex nonlinearity interdependencies of features. We regard the CF as a regression problem by the proposed model. The figure 1 shows the main structure of this distributional representation model.

### 2.1　Transforming the user and item into vectors

Each user $i \in D^{user}$ is embedding into a d-dimensional space by checking lookup table $LT_{W(.)}$ :

$$LT_{W(i)} = W_i^U$$

Where $W^U \in \mathbb{R}^{d \times D^u}$ is the parameter matrix that needs to be learned through training, $W_i^U \in \mathbb{R}^d$ is the $i^{th}$ column of $W^U$ and d is the vector size defined as hyper-parameter.

On the other hand, Each item $j \in D^{Item}$ can also be represented as d-dimensional vector by mapping the lookup table $LT_{W(.)}$:

$$LT_{W(j)} = W_j^I$$

Where $W^I \in \mathbb{R}^{d \times D^{item}}$ is the parameter matrix and $W_j^I \in \mathbb{R}^d$ is the $j^{th}$ column of $W^I$.

When the user i and item j are given as input of the recommendation system to predict the score of item j for user i , we can concatenate the user vector and item vector into a longer vector $\{W_i^U, W_j^I\}$ by applying the lookup-table to each of them.

## 2.2 Neutral Network Structure

Our proposed neutral network has three layers---an input layer, a hidden layer and an output layer. As mentioned above, the input layer is the concatenation vector $\{W_i^U, W_j^I\}$. The node of hidden layer has a full connection to the nodes in input layer and it transforms the features encoded in the user vector and item vector into real-value number by nonlinearity function. The hyperbolic tangent, or tanh, function is used as nonlinearity function as following:

$$f(z) = \tanh(z) = \frac{e^z - e^{-z}}{e^z + e^{-z}}$$

The tanh(z) function is a rescaled version of the sigmoid, and its output range is [ – 1,1] instead of [0,1]. Here z is the linear function of input vector $\{W_i^U, W_j^I\}$ and edge weight parameter $W^{L1}$ which connect the nodes between the input layer and hidden layer.

The output of the hidden layer is used as features for a logistic regression classier (the output layer) which will return the probability that means the predicted scores for item j by user i. The sigmoid function is used as the nonlinearity function which scales the output range in [0,1] and the bigger score obviously means more preference. However, the real-life applications always prefer a score in the range [0,k],say k=5. We can rescale the output of the neutral network to the right range just by multiplying result by factor k.

## 2.3 Training

We can see from the section 2.2 that the following parameters need to be

trained:
$$\theta = \{W^U, W^I, W^{L1}, W^{L2}\}$$

Here $W^{L2}$ is the edge weight of nodes between the output layer and hidden layer.

The rating records of users can be used as the training set and the training data takes the form of $[U_i, I_j, y]$ triplet. Here $y$ is the rating of user $U_i$ for the item $I_j$.

The full learning objective takes the following form of the structural risk minimization which tries to minimize the prediction error during the training:

$$\mathcal{J}(\theta) = \frac{1}{2}\sum_{i=1}^{n}(\hat{f}(\theta) - y)^2 + \lambda\|\theta\|_2^2$$

Where $\hat{f}(\theta)$ is the predicting function of distributional representation model which sequentially consists of tanh function and sigmoid function. We use standard L2 regularization of all the parameters, weighted by the hyper-parameter $\lambda$.

General back-propagation is used to train the model by taking derivatives with respect to the four groups of parameters. We use mini-batched L-BFGS for optimization which converges to a local optimum of objective function.

## 3. Experiment

### 3.1 Datasets

For evaluating our proposed model, we use the MovieLens 1M[10] and EachMovie datasets [11]. MovieLens 1M dataset contains 1000209 ratings of approximately 3900 movies made by 6040 MovieLens user and EachMovie contains 2,811,983 ratings entered by 72,916 user for 1628 different movies. For all the experiments, Ninety percentage of the rating data were randomly chosen for training and the rest 10% were used as the test set .

### 3.2 Experiment Results

RMSE is a commonly used evaluation standard for recommendation system and we use it through all experiments. In order to compare the performance of distributional representation model (DR model) with the state-of-the-art CF algorithms, we use mahout[12] as the test bed. The most commonly used recommendation algorithms such as classical KNN based model ,SlopeOne,ALS,SVD++ and improved KNN based model which was proposed by Koren[] were elaborately tuned to get as good performance as we can.

The experiment results are listed in the table 1. The best run of DR Model has the following parameters: both the length of the user vector and item vector are 24 and the number of nodes in hidden layer is 40. These results indicate consistently good performance from our DR model in both datasets and that made the distributional representation model a promising direction for further research in the collaborative filtering.

Table 1.RMSE results of MovieLens and EachMovie datasets

| Model | RMSE(MovieLens Dataset) | RMSE(EachMovie Dataset) |
| --- | --- | --- |
| User-Based KNN | 1.0476 | 0.2930 |
| Item-Based KNN | 1.0084 | 0.2618 |
| SlopeOne | 0.9370 | 0.2486 |
| ALS | 0.9550 | 0.2628 |
| SVD++ | 0.9495 | 0.2991 |
| Koren's Item-based KNN | 0.9234 | 0.2564 |
| **DR Model** | **0.9037** | **0.2409** |

## 4. Related Works

Many popular CF algorithms have been proposed in recent years. Among them, the improved item-based KNN proposed by Koren[13] and latent factor CF[14] shows great performance advantages. Latent factor CF models explain ratings by characterizing both items and users in terms of factors inferred from the pattern of ratings. One of the most successful realizations of latent factor models is based on matrix factorization[15] such as SVD and SVD++.Our proposed distributional representation model can be categorized into the latent factor CF because it explicitly encodes the latent features of users and items into the word embedding vectors. Compared with the SVD++-like matrix factorization, distributional representation model directly combine the latent factor vectors with the neutral network structure and it can explore the complex nonlinearity interdependencies of features under this framework.

As for the neutral network method or deep leaning approach in CF, RBM[16][17] and Wang's model[18] show different network structures or different optimization target compared with our proposed model.

## 5. Conclusion

We present in this paper a concise distributional representation model for collaborative filtering. To the best of our knowledge, this is the first study on the use of word embedding to the recommendation system. We can conclude from the experiment results that this model outperforms the state-of-the-art algorithms in many cases. That made the distributional representation model a promising direction for further research in collaborative filtering. If we introduce the tensor into the DR model, It's natural to regard this DR model as a special case of Tensor-based DR model. We will further explore this more general tensor-based deep leaning model in the future work.